\documentclass[apjl, iop]{emulateapj}
\usepackage{amsmath}
\usepackage{amssymb}
\usepackage{amstext}
\usepackage{apjfonts}
\usepackage{graphicx}
\usepackage{epsfig}
\usepackage{comment}
\usepackage{float}
\usepackage{color}

\newcommand{\fb}{f_{\rm B}}
\newcommand{\sig}{\sigma_0}
\newcommand{\jk}{J-K}

\newcommand{\dv}{\Delta{v}}
\newcommand{\dt}{\Delta{t}}
\newcommand{\mini}{M_{\rm ini}}
\newcommand{\Msun}{M_\odot}
\newcommand{\ptr}{P_{\rm tr}}

\shorttitle{Binarity as a function of $T_\mathrm{eff}$ and $Z$}
\shortauthors{Gao et al.}

\begin{document}

\title{The binarity of Milky Way F,G,K stars as a function of effective temperature and metallicity}

\author{Shuang Gao, Chao Liu, Xiaobin Zhang, Stephen Justham, Licai Deng and Ming Yang}
\affil{Key Laboratory of Optical Astronomy, National Astronomical Observatories, Chinese Academy of Sciences, Beijing 100012, China; \\sgao@nao.cas.cn}
 
\begin{abstract}
We estimate the fraction of F,G,K stars with close binary companions by analysing multi-epoch stellar spectra from Sloan Digital Sky Survey (SDSS) and LAMOST for radial velocity variations. We employ a
Bayesian method to infer the maximum likelihood of the fraction of binary stars with orbital periods of 1000 days or shorter, assuming a simple model distribution for a binary population with circular orbits. The overall inferred fraction of stars with such a close binary companion is $43.0\% \pm 2.0\%$ for a sample of F, G, K stars from SDSS SEGUE, and $30\% \pm 8.0\%$ in a similar sample from LAMOST. The apparent close binary fraction decreases with the stellar effective temperature. We divide the SEGUE and LEGUE data into three subsamples with different metallicity ($\mathrm{[Fe/H]} < -1.1$; $-1.1 <  \mathrm{[Fe/H]} < -0.6$; $-0.6 <  \mathrm{[Fe/H]}$), for which the inferred close binary fractions are $56\pm 5.0\%$, $56.0\pm 3\%$, and $30\pm 5.7\%$. The metal-rich stars from our sample are therefore substantially less likely to possess a close binary companion than otherwise similar stars drawn from metal-poor populations. The different ages and formation environments of the Milky Way's thin disk, thick disk and halo may contribute to explaining these observations. Alternatively metallicity may have a significant effect on the formation and/or evolution of binary stars.
\end{abstract}


\section{Introduction}
Not only are interacting binary stars responsible for numerous important stellar phenomena, but understanding the effect of binaries on stellar populations is also important for correctly interpreting the integrated light of stellar populations \citep[see, e.g.,][]{eldridge2008} and the internal structure of the Milky Way \citep[see, e.g.,][]{juric2008}. Natural differences in binary-star features may also have been misinterpreted as variations in the stellar initial mass function \citep{maccarone2014}. The initial properties of binary stars are also a clue to understanding the star formation process \citep[see, e.g.,][]{duchene2013}.

However, our knowledge of stellar binarity is far from perfect. Regulus ($\alpha$ Leonis) is one of the brightest stars in the sky, but was only recently discovered to be in a binary system with a $\approx$ 40 day orbital period \citep{gies2008}. Given that the nature of this extremely well-studied star was unknown for so long, our detailed knowledge of stellar binarity in general seems likely to be highly incomplete. 

Prior studies have nonetheless convincingly found that being single is the exception rather than the norm for massive stars \citep[see, e.g.,][]{,abt1976,abt1978,kobulnicky2007,eggleton2008,sana2012}. For example, \citet{sana2012} concluded that less than a third of O-type primary stars will evolve as if they were single.   \citet{eggleton2008} concluded that the average multiplicity of O-type stars is \emph{more than} 2, i.e.\ even binaries have \emph{lower} multiplicity than the expectation value for an O-star.

It is generally accepted that the binary fraction drops with decreasing stellar mass \citep[see, e.g.,][]{lada2006,eggleton2008,clark2012,raghavan2010}. However, that measures of the present-day binarity in samples of later spectral-type stars are only an indirect measure of the initial binarity of those stars. Inferring the initial properties from the present-day population is a hard problem, given the many ways in which binary systems can evolve. There is some indication that apparent binarity decreases with age \citep[see, e.g.,][and references therein]{duchene2013}, and the distribution of binary periods and mass-ratios should also be expected to change as binaries interact through their lifetimes. 

Despite the results described above, our knowledge of stellar binarity still contains many unknowns \citep[for a review see][]{duchene2013}.  

Large spectroscopic surveys provide a powerful way to investigate the binarity of Galactic stars, since multi-epoch spectroscopy allows binaries to be detected via radial velocity (RV) variations. The Sloan Digital Sky Survey (SDSS) has already been used for this purpose \citep[see, e.g.,][]{york2012,badenes2012,maoz2012,herringer2013}.  Over the next few years the Guoshoujing telescope \citep[also known as LAMOST, see][for the overview]{,cui2012,zhao2012} will survey the stellar population of the Galaxy, providing spectra for more than five millions Galactic stars, including many with multiple observations.  

This work was begun to investigate the power of the existing LAMOST and SDSS observations to examine stellar binarity. The first official LAMOST data release (DR1) contains data from both the pilot \citep{luo2012} and regular surveys until 2013 June. It contains one million stellar spectra and their derived physical parameters (RV, effective temperature $T_\mathrm{eff}$, surface gravity $\log{g}$, and metallicity $\mathrm{[Fe/H]}$). The observed and released data include Galactic anti-center area \citep{liu2013}. We also employ data from the SDSS DR9 \citep{ahn2012}, which includes spectra taken over 14,555 deg$^2$ of the sky and data from the SEGUE project.

Binaries complicate the use of stars as tracers of galaxy evolution, and galaxy structure and evolution may in turn complicate the interpretation of observed differences in stellar populations. In particular, when comparing Galactic samples of present-day stars with different metallicities, it may be significant that the different-metallicity samples are drawn from different parts of the Milky Way's structure. For example, different formation scenarios for the Galactic thin and thick disks might affect the present-day binarity of the stars in those populations as much as the differences in age and metallicity \citep[see, e.g.,][]{gilmore1983,liu2012}. 

To simplify the language, this work generally refers to a ``binary fraction'' (denoted $\fb$). However, this only represents the fraction of primary stars which have a binary companion with an orbital period that leads to RV variations which we can detect, not the overall fraction of stellar systems which are binaries. This is also not the same as the fraction of stars which are in a binary system. 

Our analysis method is presented in Section \ref{sec:method}. The SDSS and LAMOST samples and our analysis are described in Section \ref{sec:data}, with our conclusions discussed in Section \ref{sec:conclusions}.

\section{Method}
\label{sec:method}

Binaries can be detected by comparing the RVs obtained from two epochs for the same stars. However, RV differences may also occur simply due to uncertainty in the two measurements. Therefore, the probability of the difference of RVs from two epochs, $\Delta v$, for a group of stars is contributed by both the motion within binaries and the uncertainty of the velocity measurement:
\begin{equation}\label{eq1}
p(\dv)=\fb p_\mathrm{B}(\dv|\sig,\dt,\mathcal{M_B})+(1-\fb)p_S(\dv|\sig),
\end{equation}
where $\fb$ is the binary fraction of the sample of stars, and the mean measurement error is $\sig$. The probability of observing $\dv$ for binary stars is $p_\mathrm{B}(\dv|\sig,\dt,\mathcal{M_B})$, given a time separation $\dt$ between the two observations and a binary dynamical model $\mathcal{M_B}$. The probability of observing a given $\dv$ for a single star, $p_\mathrm{S}(\dv|\sig)$, depends only on the measurement error.

Clearly $p_\mathrm{B}$ depends on the assumed dynamical model for the binary population. We assume that: (1) the observed stars with companions are on the main sequence, with masses drawn from the Salpeter IMF \citep{salpeter1955}; (2) the mass ratio\footnote{The mass ratio is defined as the ratio of the mass of the secondary star to that of the primary star. Note that our model assumes that the observed star is the most massive star in the binary, which may be incorrect if the secondary star is compact. However, the fraction of neutron-star and black-hole secondaries should be negligible.} follows a uniform distribution between 0.05 and 1; (3) the orbit of each binary is circular ($e=0$), with the orbital periods following the log-normal distribution given by \citet{raghavan2010}; (4) we consider random orientation of the systems in 3D space and for the initial orbital phases $\phi_0$.

With the above assumptions, a random $\dv$ can be calculated from the model by drawing an orbital period, the stellar mass of the primary, a mass ratio, an orientation, and an initial phase from their corresponding distributions given a fixed $\dt$. For a sample of stars with two-epoch RV measurements, one can then estimate $\fb$ and $\sig$ using a maximum-likelihood method based on Equation \ref{eq1}.

\begin{figure}
\begin{center}
  \plotone{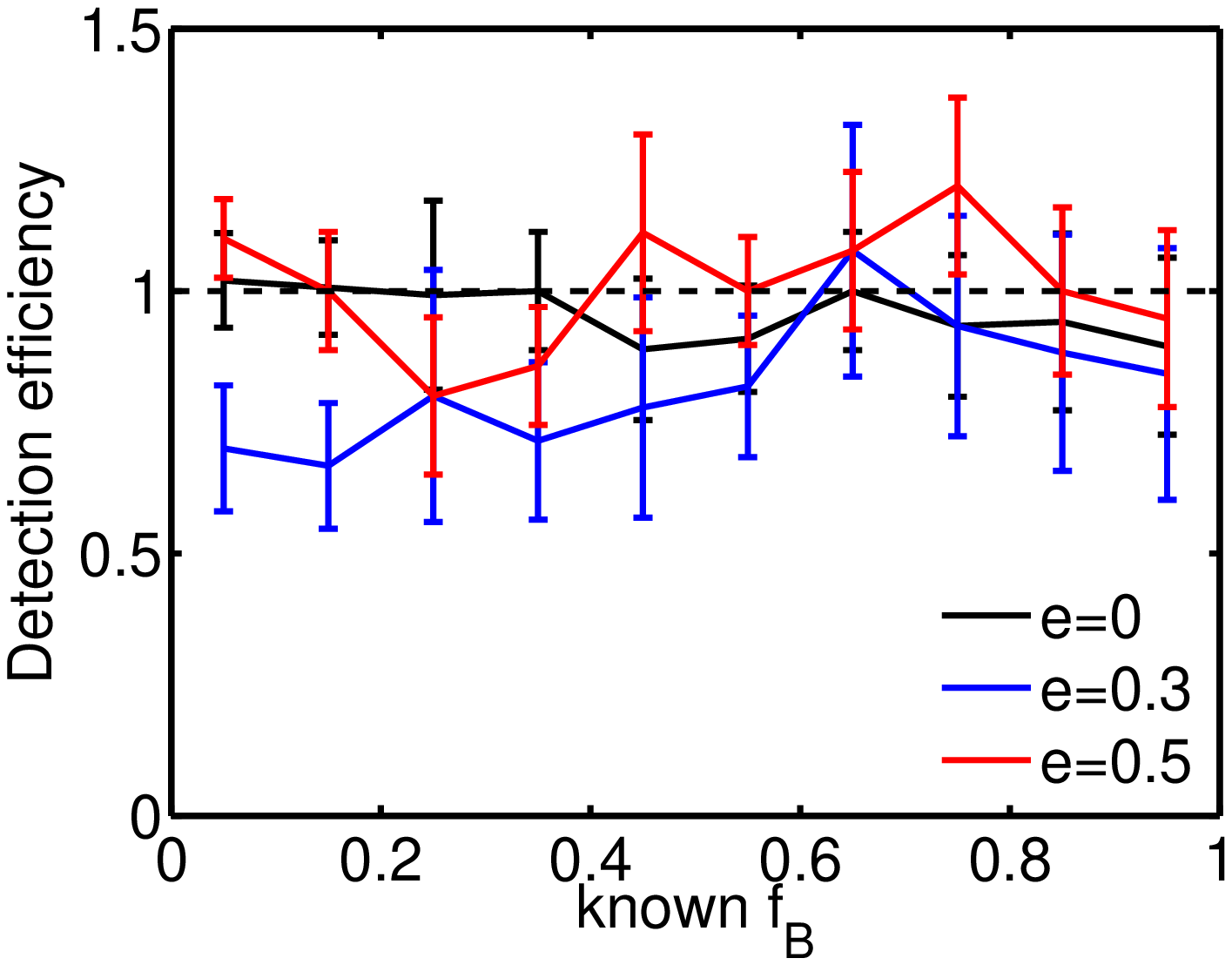}
  \plotone{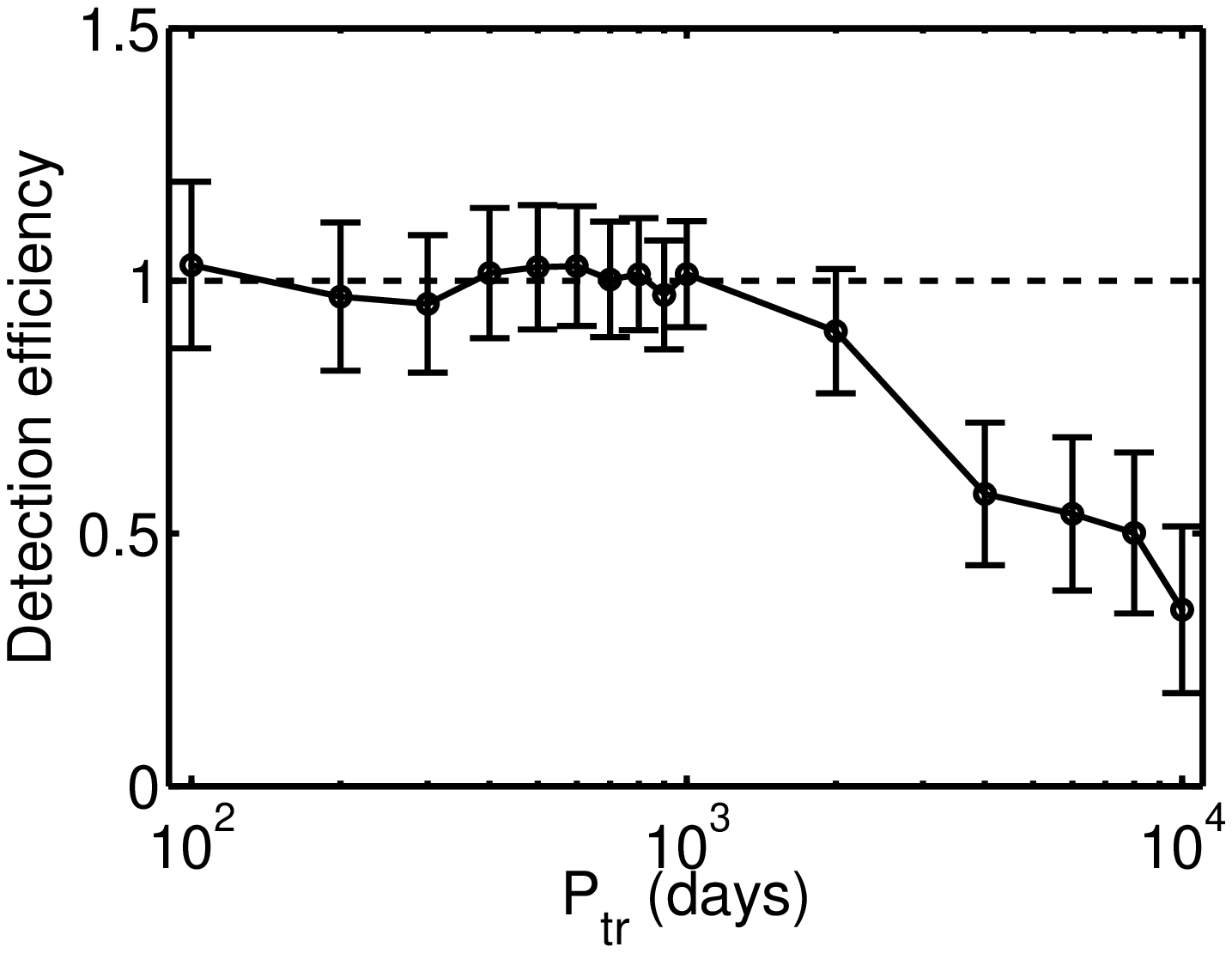}
\end{center}
\caption{We show the detection efficiency, i.e.\,the ratio of the inferred $\fb$ to the known $\fb$ in the artificial input sample, as a test of our method. Uncertainty estimates are from the marginalized likelihood distribution. Upper panel: input samples with different fixed eccentricities ($e=0, 0.3, 0.5$) do not show a single overall systematic bias over the range of potential $\fb$, although we tend to underestimate the binary fraction for $e=0.3$. In almost all cases the error bars overlap. The change in detection efficiency with $\fb$ results from the covariance between the inferred $\fb$ and $\sig$ (see Figure \ref{fb}). Lower panel: detection efficiencies for different $\ptr$. This indicates that we can reliably infer $\fb$ for periods up to 1000 days.}
\label{sim}
\end{figure}

Since we know that our assumption of zero eccentricity is wrong for real binaries, we tested how much non-zero eccentricities affect our determination of $\fb$. Figure \ref{sim} shows that our ability to recover the true $\fb$ for artificial samples with $e=0.3$ and $0.5$ is worse than for $e=0$, although the results are generally consistent. The largest potential bias this reveals is that, with $e=0.3$, we underestimate the binary fraction for $\fb \lessapprox 0.5$.  While we consider that this effect does not affect our qualitative conclusions, the additional uncertainty should not be forgotten.

In practice, we cannot detect binaries for which the orbital period is so long that the velocity difference over time $\dt$ is dominated by the measurement error in the velocities. This effectively truncates the orbital period distribution at $\ptr$, which is a parameter we must estimate. The mock sample is created with a known $\fb$, i.e.,\,80\% and maximum period of $10^4$ days. Figure \ref{sim} shows results using simulated data, the real $\dt$ distribution for our LAMOST sample and $\rm \sig = 4\,km~s^{-1}$. (For both LAMOST and SDSS, $\dt$ is typically less than 300 days and the RV measurement error is $ \rm \approx 4\,km~s^{-1}$.)  Based on these calculations we chose $\ptr = 1000$ days, above which our detection efficiency begins to decrease.

\section{Data and Analysis}
\label{sec:data}

\subsection{LAMOST}
We excluded data with $\dt<1$ day and Galactic latitude $|b|<20^\circ$ due to extinction. The time gap distribution is shown as the left upper panel of Figure \ref{fb}. The sample was cross matched with $JHK$ photometry from the 2MASS catalog \citep{skrutskie2006}, and extinction correction was performed using the Schlegel dust map \citep{schlegel1998}. We select F/G dwarf stars using the condition $0.2<(\jk)_0<0.45$, such that the distance and the mass of each star could be obtained from the pipeline. From this procedure, we obtained a sample of 5204 F/G stars with $\dv$ measurements.

\begin{figure}
\begin{center}
    \plotone{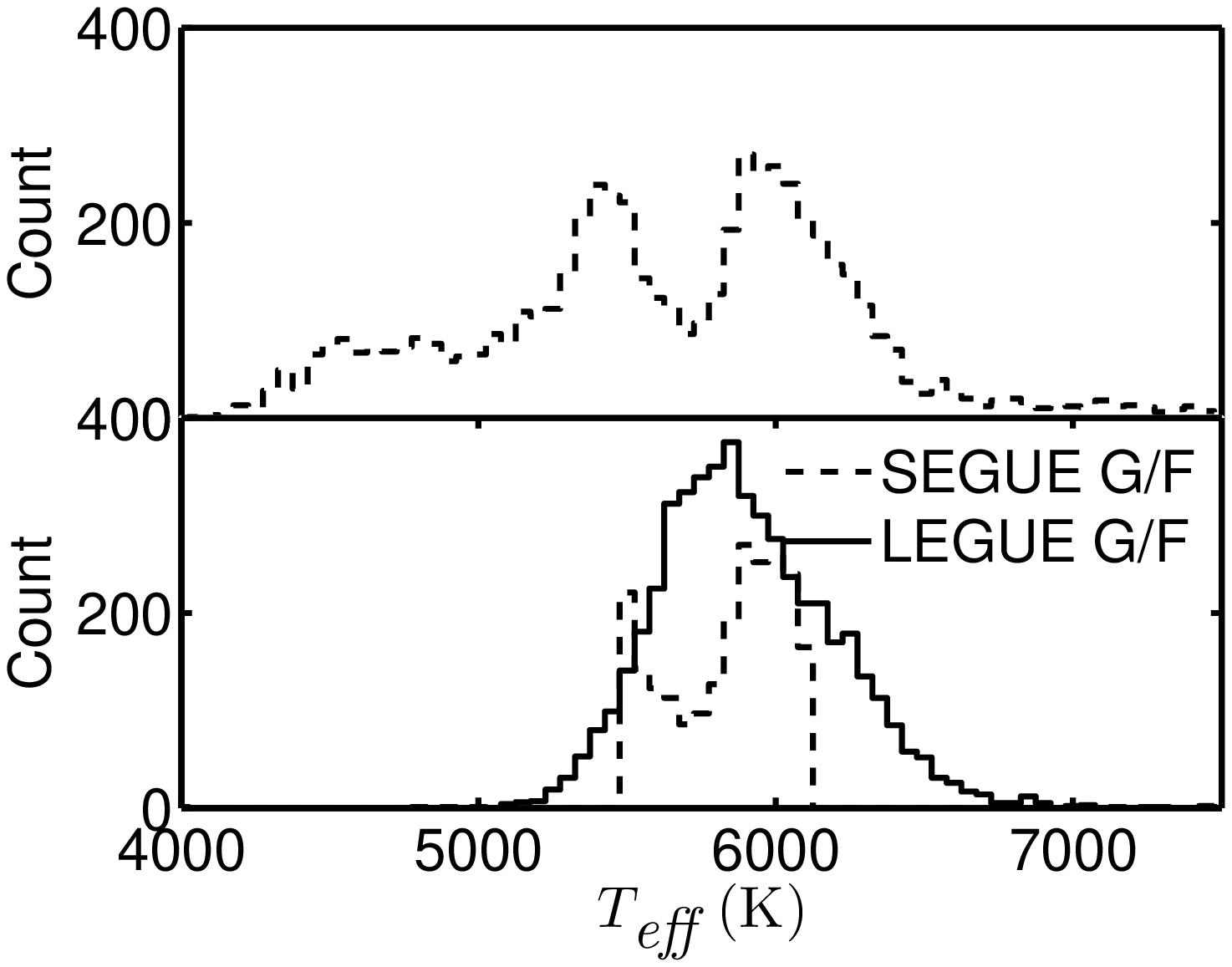}
     \plotone{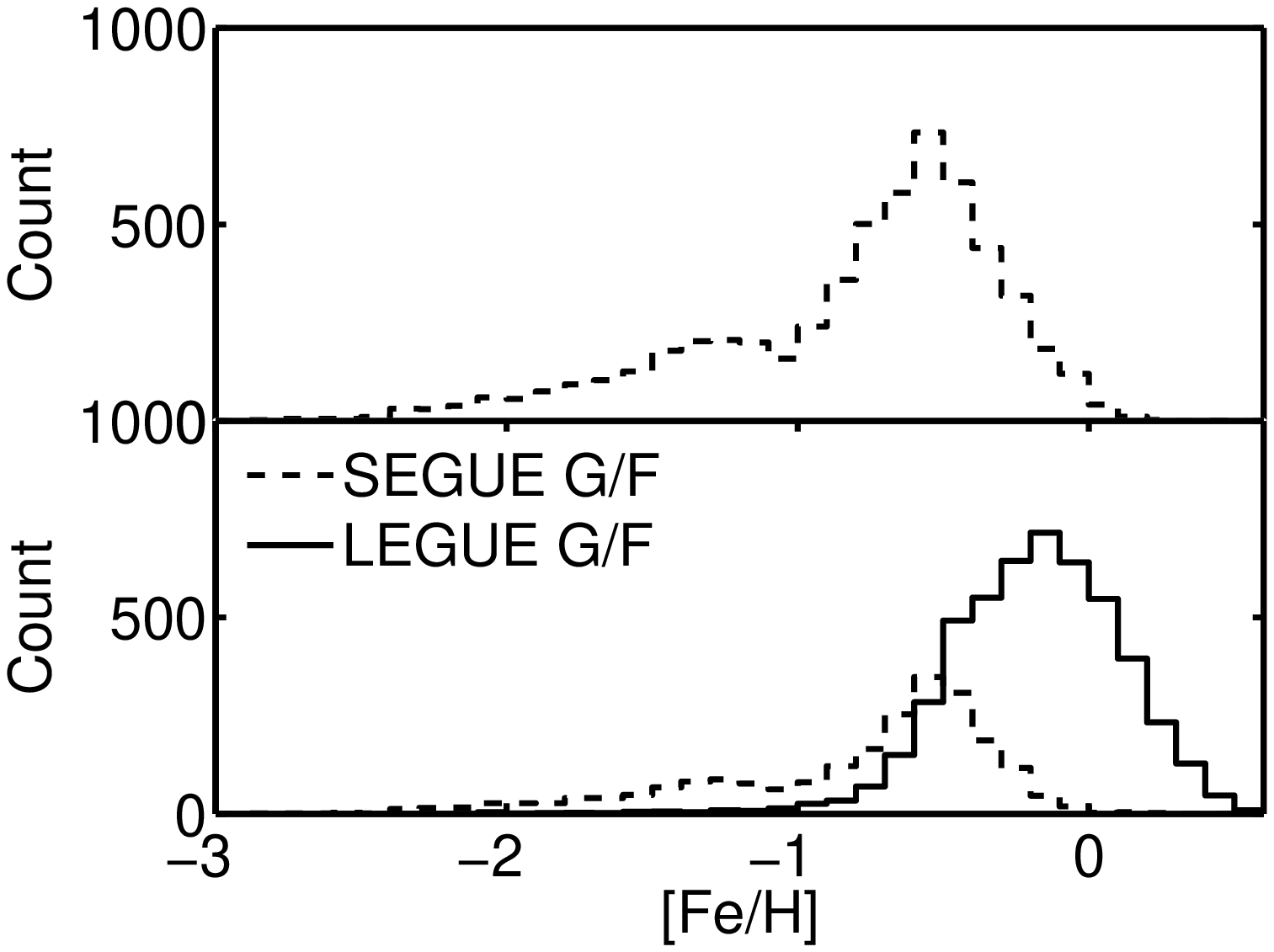}
     \plotone{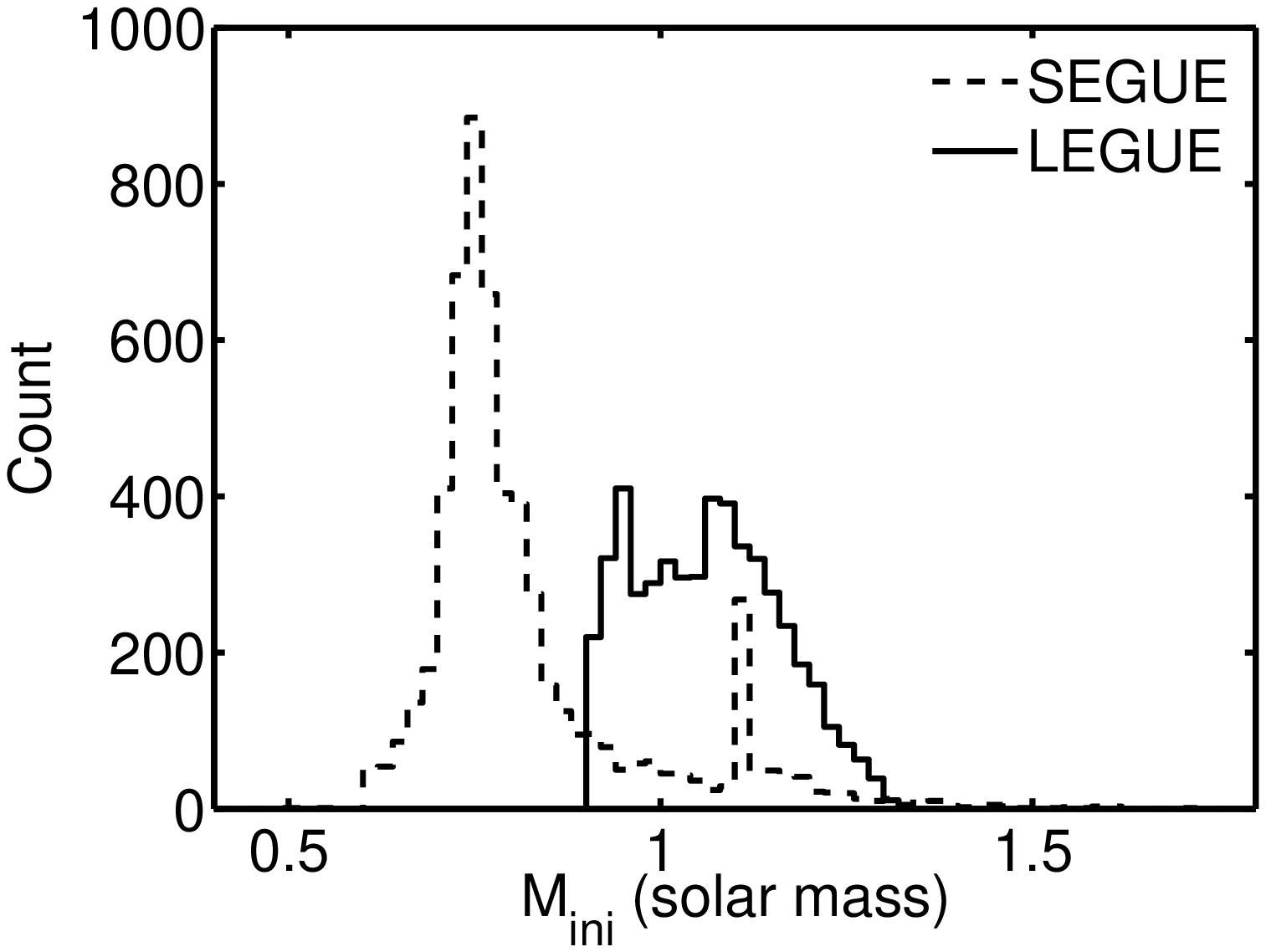}
     \plotone{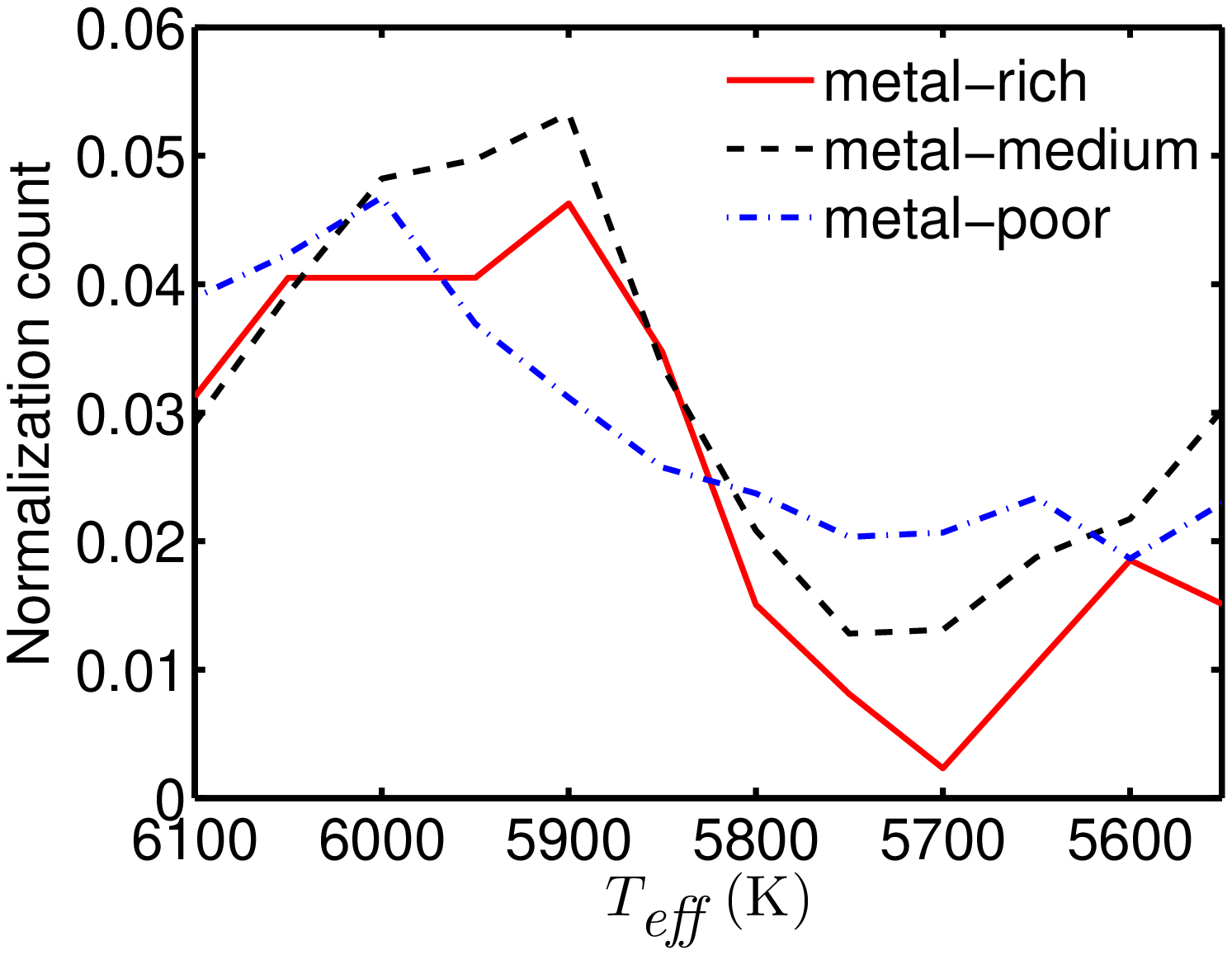}\\
\end{center}
\caption{Distribution of $T_\mathrm{eff}$ (the first panel),  $\mathrm{[Fe/H]}$ (the second panel) and mass (the third panel), and selected $T_\mathrm{eff}$ distributions of three $\mathrm{[Fe/H]}$ bins (the last panel) of two samples. In the first two panels, the upper subfigures show the global sample of SDSS and the lower ones show the selected F/G-type stellar sample from SEGUE and LEGUE. The last panel shows all three $T_\mathrm{eff}$ distributions with different $\mathrm{[Fe/H]}$ of selected F/G stars from SEGUE.}
\label{dis}
\end{figure} 

The parameters distributions of sample are shown in Figure \ref{dis}. The weight $w_i$ is defined following the method of \cite{liu2012}. Using methods described in Section 2, the LAMOST data produced the results shown in Figure \ref{fb}. From LAMOST sample, we infer that $30\%$ of F/G primary stars have a binary companion and an orbital period in the range to which we are sensitive. The marginalized distributions are used to estimate the dispersions of parameter estimations in results for $\fb$ and $\sig$, giving $\fb=30\%\pm8\%$ and error $\sig=4.5\pm0.18$ $\rm km s^{-1}$.

\begin{figure}
\begin{center}
 \plottwo{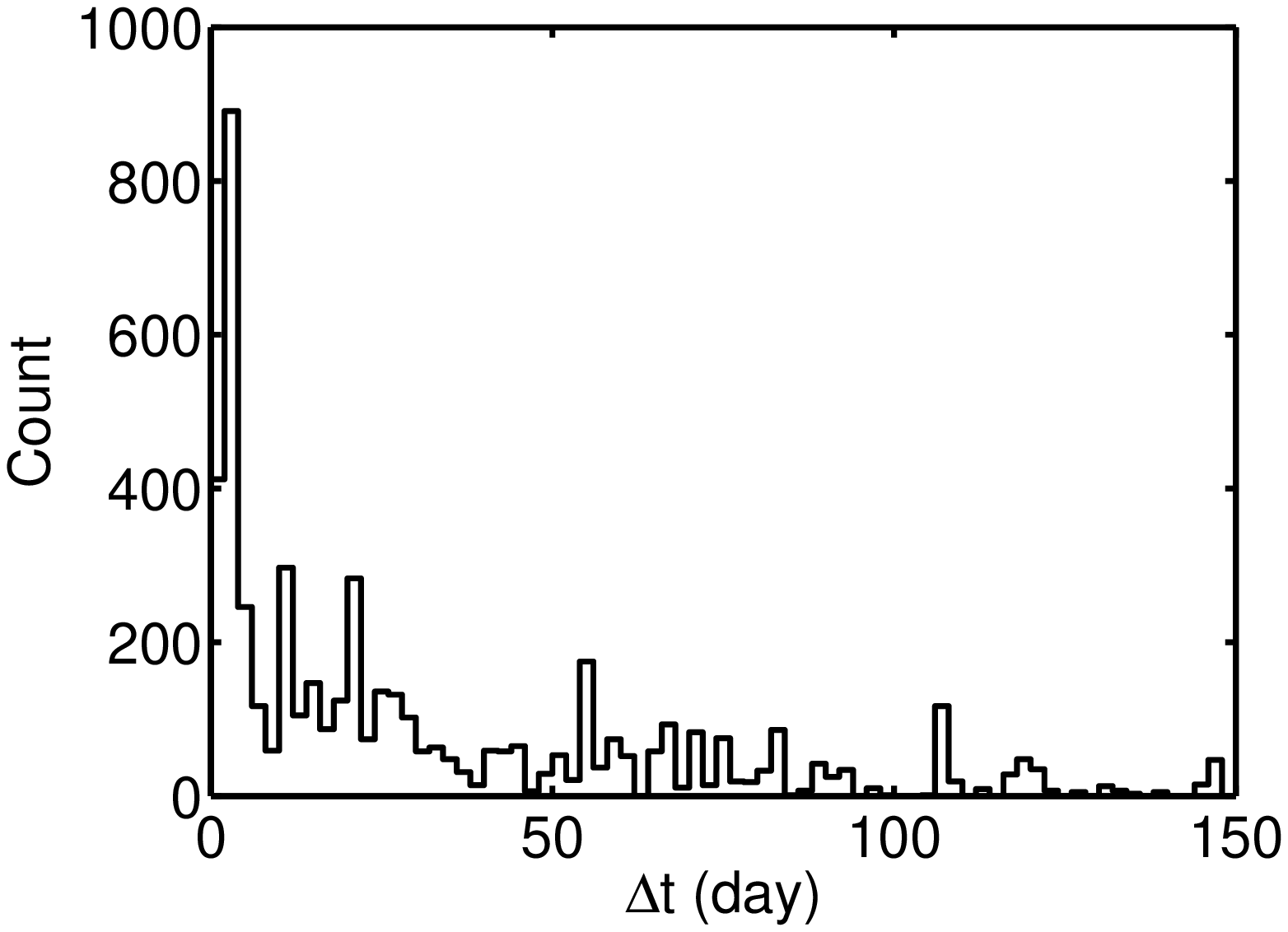}{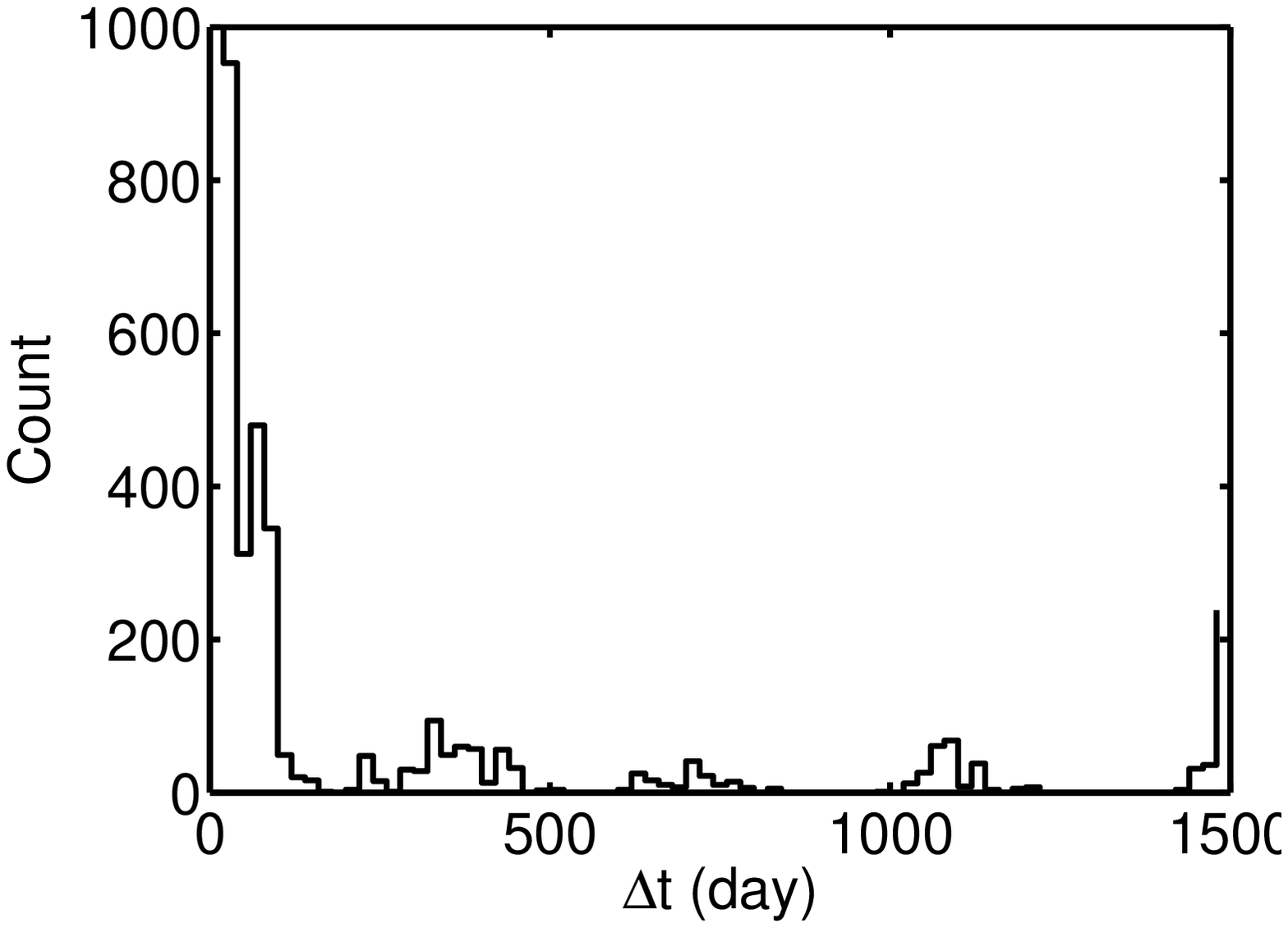}
 \plottwo{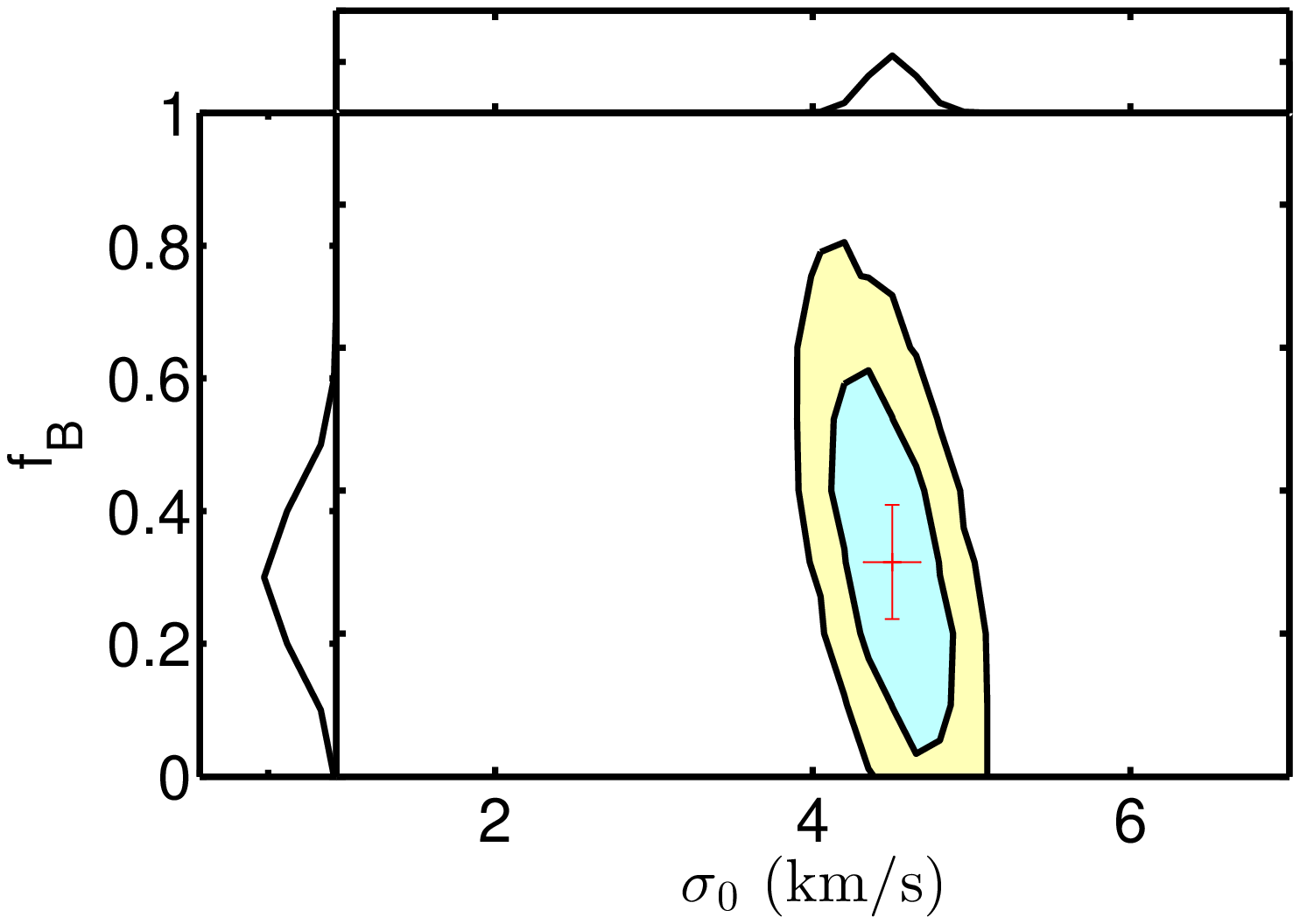}{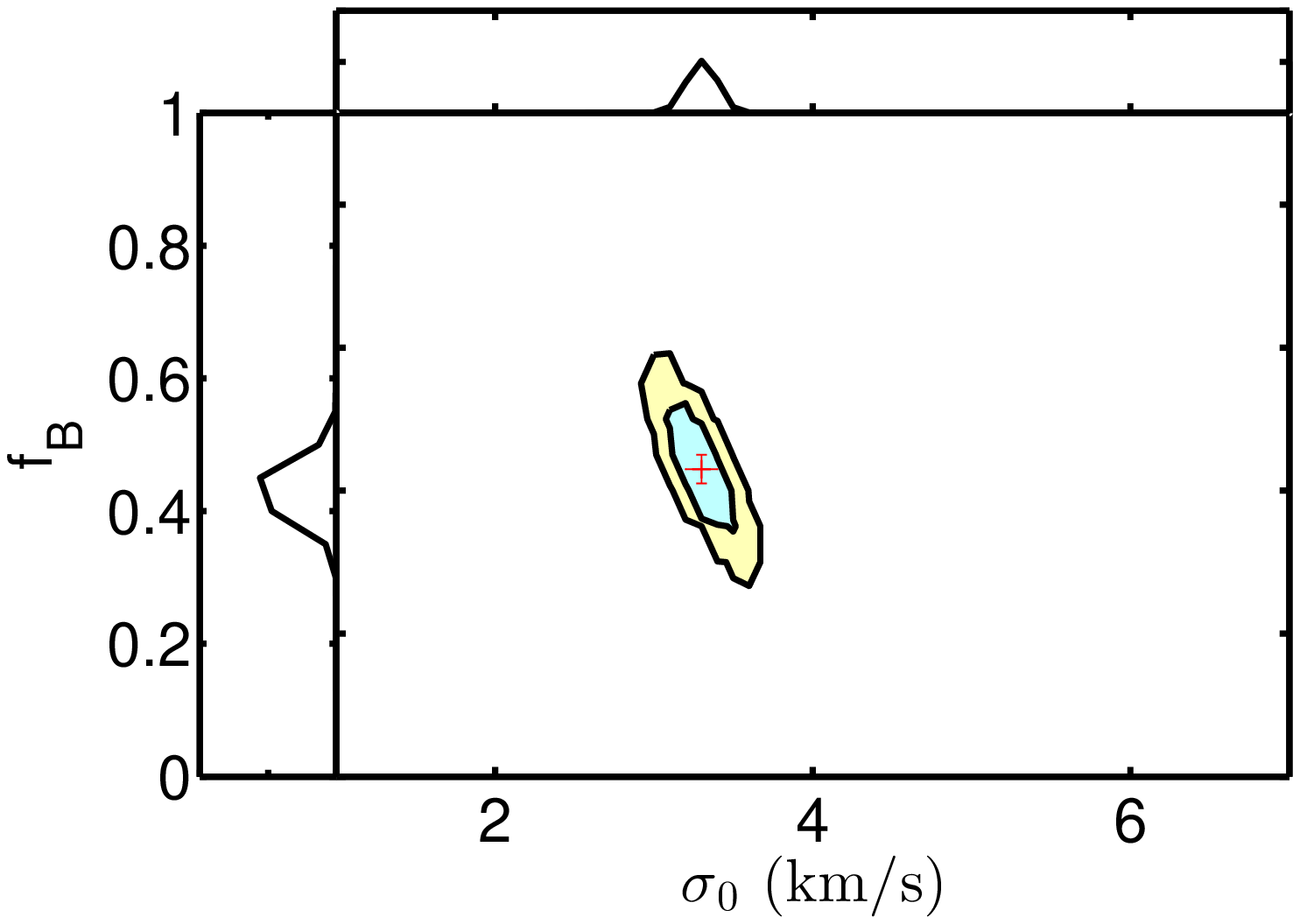}
  \plottwo{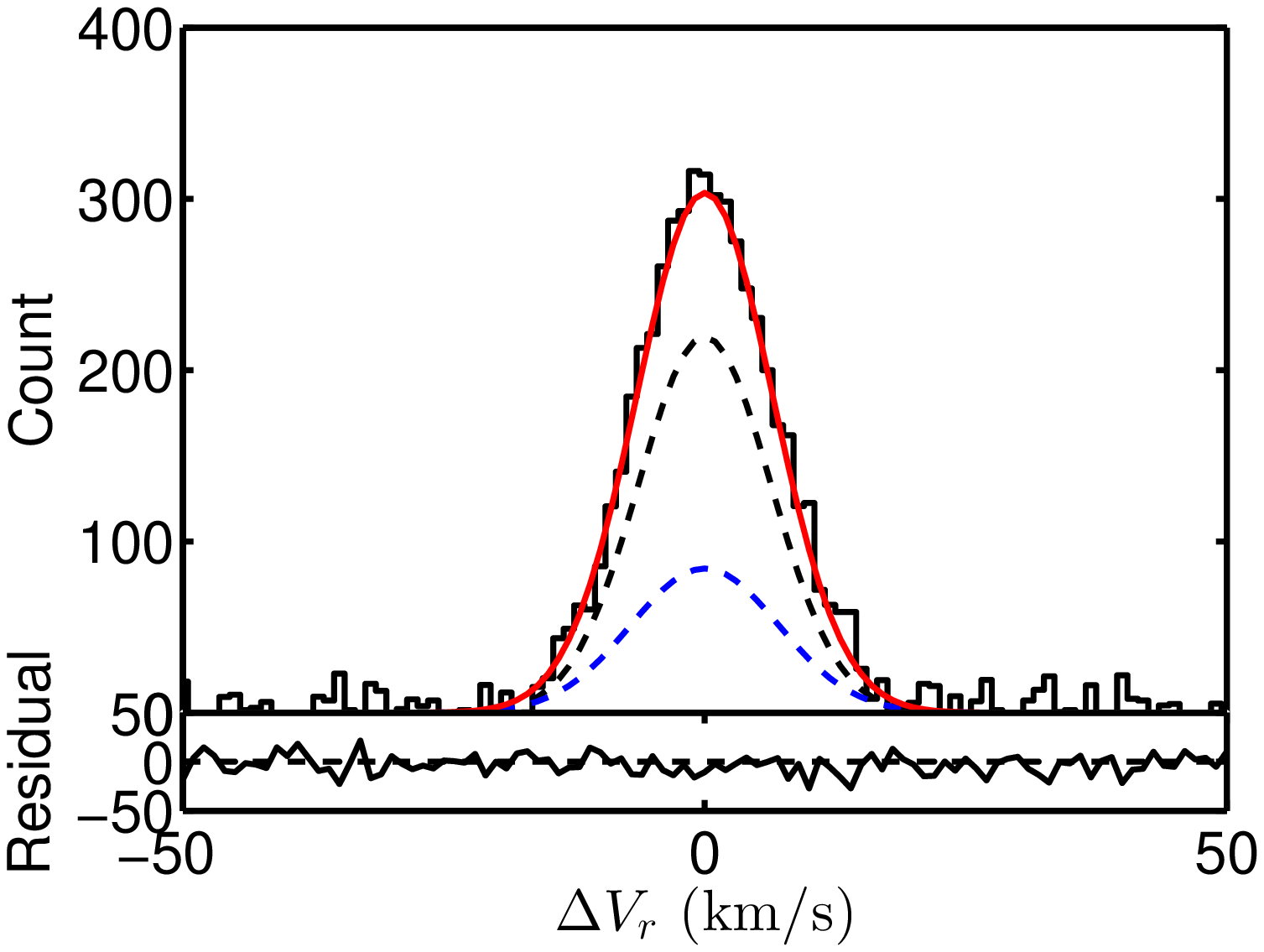}{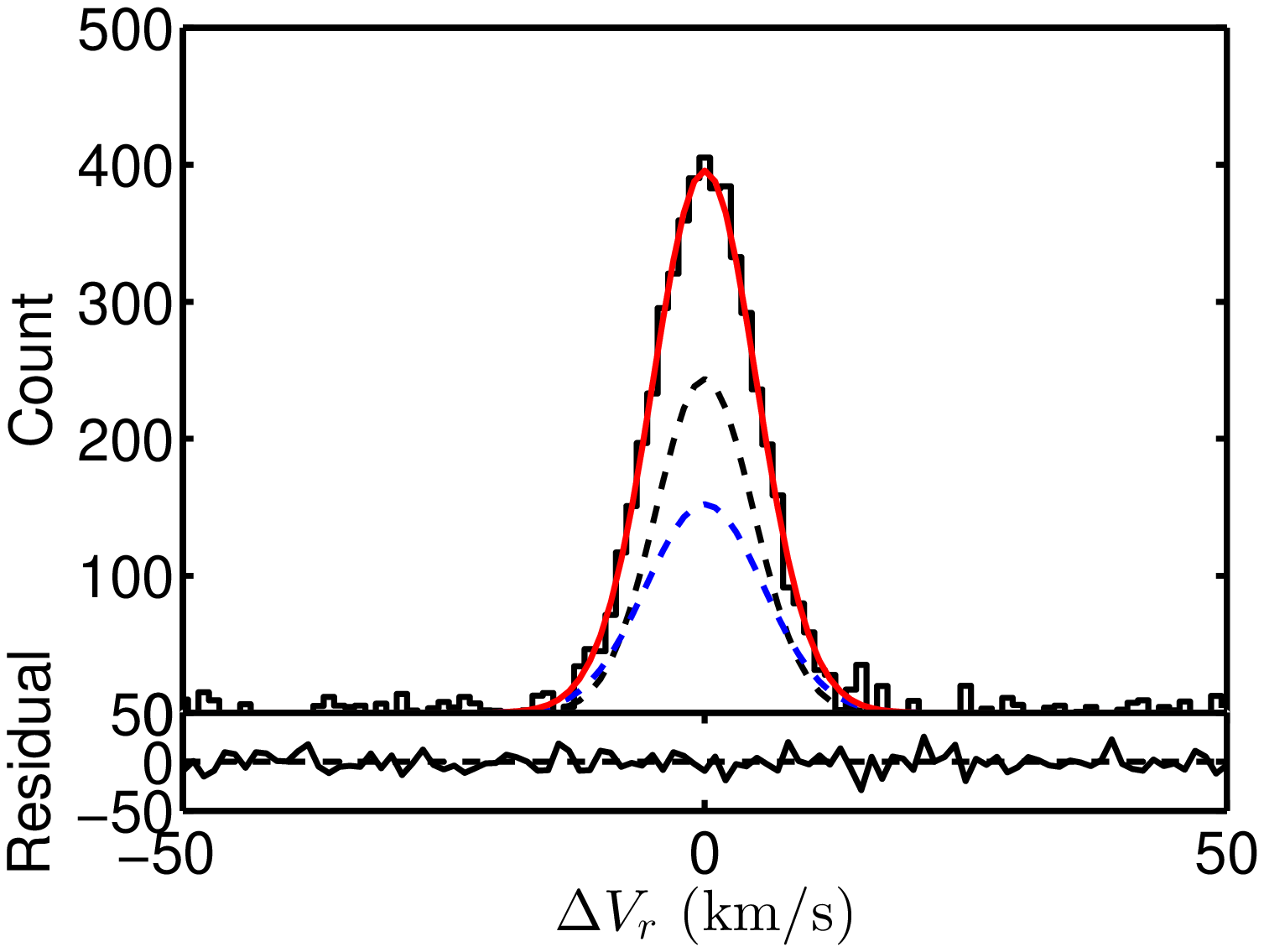}
\end{center}
\caption{Left panels shows LEGUE information, the right column SEGUE. Top panels: the distributions of time gaps between observations of the same star in the two samples. Middle panels: maps of the posterior PDFs in the plane of $\sig$ and $\fb$. The shaded contours contain $\sigma$ and $3\sigma$ of the cumulative PDF, with marginalized 1-d PDFs displayed at the edges of the plane. Bottom panels: the dashed curves show model RV differences for the single and binary components, with the (red) solid curves the combined population. The histograms are the bias-corrected observed profiles.}
\label{fb}
\end{figure} 

\subsection{SDSS}
We select 5,728 targets with twice observations from SDSS DR9 SEGUE catalog \citep{ahn2012,lee2008a,lee2008b}. The statistical weights given to each star are determined in the same way as for the LAMOST sample. 

The $M_\mathrm{ini}$ were estimated by comparing $T_\mathrm{eff}$, $\log{g}$, and $\mathrm{[Fe/H]}$ of each star with Girardi's isochrones \citep{giradi2000}. We exclude targets with $T_\mathrm{eff}>7500$ K and only include stars with $\mini$ between 0.6 and 1.2 $\Msun$. Figure \ref{fb} shows results after the stellar masses, $T_\mathrm{eff}$, color indices, and weights are considered to corrected for incompleteness. Uncertainties are again estimated using the marginalized distributions. From this SEGUE sample, we infer $f_{B} = 43.0\% \pm 2.0\%$ and $\sig=3.3 \pm 0.10$ $\rm km s^{-1}$.

\subsection{Binary Fractions as a Function of Spectral Type}

We divided the primary stars from the LEGUE and SEGUE samples into sub-samples based on $T_\mathrm{eff}$. For each bin of $T_\mathrm{eff}$ we re-calculated $\fb$. Between 4000 and 7500 K, $\fb$ changes significantly, as shown in Figure \ref{teffbin}. The inferred RV errors vary from 2 to 5 $\rm km s^{-1}$. The RV errors change with the stellar spectral types due to the changing availability of spectral lines. The stars with a $T_\mathrm{eff}$ of $\sim6000$ K have the best RV measurements. Three sub-samples from our selected LAMOST stars are shown in Figure \ref{teffbin}. As shown in Figure \ref{teffbin}, we find a higher value of $\fb$ for hotter stars. This trend has previously been identified \citep[see, e.g.,][]{eggleton2008,raghavan2010}.

\subsection{Binary Fractions as a Function of Metallicity}

\begin{figure*}
\begin{center}
    \plotone{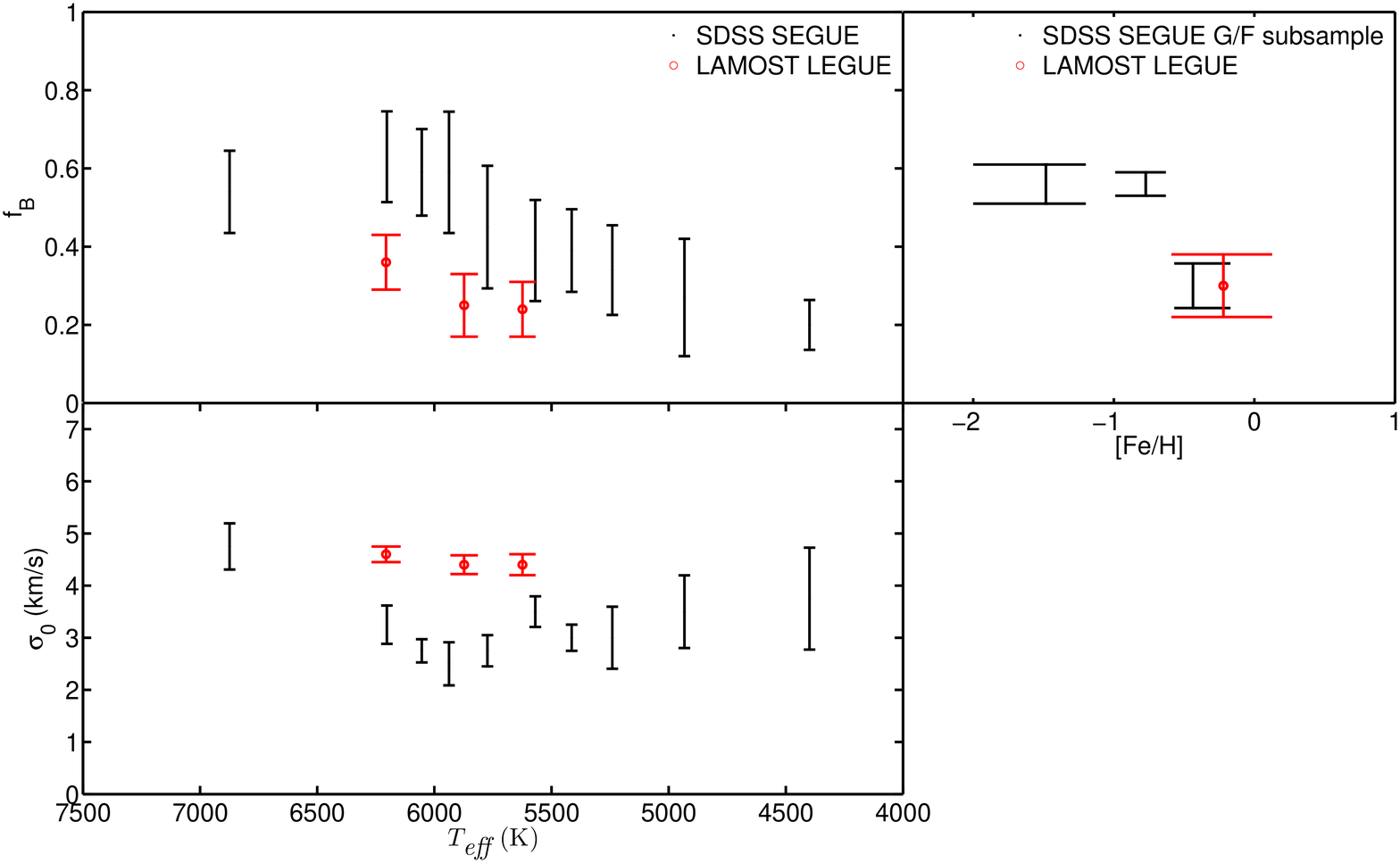}
\end{center}
\caption{Binary fraction as the function of $T_\mathrm{eff}$ and $\mathrm{[Fe/H]}$. The binary fraction and RV errors are limited within 10 stellar $T_\mathrm{eff}$ bins for SEGUE sample, which are shown as black error bars in the upper and lower panels. For LAMOST sample, result based on three bins are shown as red color in panels together. The $\mathrm{[Fe/H]}$ function are shown in right panel.}
\label{teffbin}
\end{figure*}

We also selected three sub-samples with different $\mathrm{[Fe/H]}$ from the SEGUE sample of F/G stars (i.e.,\ over a limited range of $T_\mathrm{eff}$, as shown in Figure \ref{dis}), and then re-calculated Figure \ref{fb} for those $\mathrm{[Fe/H]}$ groups (with metal-poor defined such that $\mathrm{[Fe/H]}<-1.1$; moderate-metallicity with $-1.1<\mathrm{[Fe/H]}<-0.6$; and metal-rich stars with $\mathrm{[Fe/H]}>-0.6$). The results are shown in Figure \ref{teffbin} and Table \ref{result}, displaying a clear change in $\fb$ with $\mathrm{[Fe/H]}$. 

One possibility was that this result might have been an artifact of sample selection, e.g., because the different $\mathrm{[Fe/H]}$ samples contain different fractions from each spectral type. Hence, we calculated normalized star count in each spectral type bin for each $\mathrm{[Fe/H]}$ subsample (as shown in the last panel of Figure 2). These star counts display similar changes with spectral type, which does not show the apparent $\mathrm{[Fe/H]}$ effect.  Hence this systematic change in $\fb$ with $\mathrm{[Fe/H]}$ is not a result of the change with spectral type combined with sample selection, but a genuine separate systematic effect.

\begin{deluxetable*}{lrcccc}
\tabletypesize{\scriptsize}
\tablewidth{0pt}
\tablecaption{Final Results of $\fb$ and $\sig$ among Two Catalogs\label{result}}
\tablehead{\colhead{Sample} & \colhead{Used count} & \colhead{$\fb$} & \colhead{$\sigma(\fb)$} & \colhead{$\sig$ ($\rm km s^{-1}$)} & \colhead{$\sigma(\sig)$ ($\rm km s^{-1}$)}
}
\startdata
SDSS SEGUE all & 5728  & $43\%$  & $2.0\%$  & 3.3  & 0.10  \\
~~~~~Metal-poor F/G & 607  & $56\%$  & $5.0\%$  & 5.6  & 0.16  \\
~~~~~Metal-medium F/G & 808  & $56\%$  & $3.0\%$  & 3.2  & 0.16  \\
~~~~~Metal-rich F/G & 1,134  & $30\%$  & $5.7\%$  & 3.4  & 0.15  \\
LAMOST LEGUE F/G & 5204  & $30\%$  & $8.0\%$  & 4.5  &  0.18 
\enddata
\tablecomments{The two $\sigma()$ operators denote the dispersions of $\fb$ and $\sig$ estimations.}
\end{deluxetable*}

\section{Discussion and Conclusions}
\label{sec:conclusions}

We have inferred the fraction, $\fb$, of stars with a binary companion in an orbit with a period less than roughly 1000 days. Our results reproduce the qualitatively expected decrease in binary fraction with decreasing $T_\mathrm{eff}$ of the primary star. At the same time, we have estimated the intrinsic RV error for LAMOST. Our result ($\rm 4.5~km~s^{-1}$) is somewhat smaller than the ``error'' given in the LAMOST DR1 catalogue, which was an obvious overestimate. The LAMOST technical department are currently preparing more reliable error determinations.

The absolute values we infer for $\fb$ are very likely sensitive to whether our assumed period distribution accurately represents the real period distribution in the period range to which we are sensitive. We have also simply assumed that the binary orbits are circular, as discussed and tested in Section \ref{sec:method}, which may well limit our accuracy. When more RV epochs are available for more systems then we should be able to constrain the form of the eccentricity and period distributions whilst finding $\fb$ (as achieved for massive stars by \citealt{sana2012}), rather than assuming a distribution. Nonetheless, the relative changes which we find in $\fb$ do indicate that either $\fb$ or the properties of the orbital distributions are changing as a function of spectral type.

Our most striking result is the change of $\fb$ with $\mathrm{[Fe/H]}$. For the SDSS SEGUE samples containing metal-poor and metal-medium stars we find a substantially higher value of $\fb$ ($56\%$) than in the metal-rich sample ($30\%$). To our knowledge, this is the first time that such a strong metallicity-dependence in binary populations has been detected, whatever the effect is purely due to a systematic change in $\fb$ or to some other aspect of the orbital distributions.

This observational difference has several possible interpretations, partly since we are not sensitive to the overall binary fraction but only to binaries with present-day orbital periods less than 1000 days. Furthermore, the population ages are very likely to be different between the samples. Assuming that the metal-rich sample contains the youngest stars, potential explanations include:

\begin{itemize}
\item The formation of metal-rich stars might generally lead to an intrinsically lower binary fraction, or to systematically wider binaries. The stellar IMF may be built up as a consequence of dynamical interactions between protostars (for which see, e.g., \citealt{price1995}); some of these models predict that binaries are more commonly produced in gas-rich environments \citep{bate2002}. If the gas fraction during the star formation epoch decreases with increasing $\mathrm{[Fe/H]}$, which seems plausible, then this prediction would be consistent with our results. If so, then we have detected an imprint of the star formation process.

\item The samples might have had identical initial binary fractions. Over time, interactions may harden some of the initially wide binaries, reducing their separation and orbital period and therefore increasing the fraction of the binaries which we can detect. Assuming the high $\mathrm{[Fe/H]}$ sample is the younger one, the lower $\mathrm{[Fe/H]}$ samples would have had longer for such interactions to occur.

\item Even if the samples had identical initial binary fractions \emph{and} ages, the different $\mathrm{[Fe/H]}$ might have led to different present-day period distributions, i.e., $\mathrm{[Fe/H]}$ might alter the outcome of the binary interactions \citep[see, e.g.,][]{ivanova2006}. 

\item In the older samples, many of the primary stars we detect could once have been secondary stars, i.e.,\ were companions to more massive stars which have now become a low-luminosity, compact remnant. Those F/G/K secondary stars would not be detected as F/G/K stars in the younger population, and would hence not be included in our sample, since the light from the system would be dominated by the more massive primary.

\item The specific formation environments of these particular Milky Way stellar populations may have led to systematically different binary fractions or binary period distributions as a function of $\mathrm{[Fe/H]}$. This would suggest that the type of star formation which led to the Milky Way thick disk and halo is different to the style of star formation which produces thin disk stars. In turn this would indicate that the thick disk has not been formed by radial migration of stars from the thin disk. Other argument have previously been used to suggest that the thick disk was not formed by migration of thin disk stars \citep{liu2012}, so we note that the chain of logic is not reversible (i.e.,\ this conclusion about the structure of the Galaxy would not imply that this is the correct explanation for the apparent difference in binary fractions). 
\end{itemize}

Clearly more than one of these statements might simultaneously be significant in explaining the difference in binary fraction which we have inferred.

The LAMOST sample contains data from stars which were observed in the nearby Galactic thin disk, i.e. stars with $\mathrm{[Fe/H]}$ similar to those of the metal-rich stars selected from the SDSS sample ($\mathrm{[Fe/H]}>-0.6$). Hence the $\fb$ inferred from LAMOST DR1 is consistent with the SDSS result. Future LAMOST observations should enable us to study the binary fractions in more detail. We suggest that the future LAMOST selection function should be considered with this aim in mind. Future measurements of $\mathrm{[\alpha/Fe]}$ for the stars would also be helpful.

Moreover --- \emph{whatever} the origin of this population variation --- studies of internal galaxy kinematics, galaxy structure and evolution should be careful not to assume that the close binary fraction does not change as a systematic function of $\mathrm{[Fe/H]}$.

\acknowledgments
This work is supported by the National Key Basic Research Program of China 2014CB845703/4, the Strategic Priority Research Program ``The Emergence of Cosmological Structures'' of the Chinese Academy of Sciences, Grant No. XDB09000000, and the Young Researcher Grant of National Astronomical Observatories, Chinese Academy of Sciences (NAOC). SG and MY are funded byLAMOST fellowship of NAOC. CL acknowledges the National Science Foundation of China (NSFC) grant 11373032 and U1231119. XZ acknowledges Project 11373037 supported by NSFC. SJ acknowledges support from NSFC grant 11250110055 and 11350110324. SG thanks Xiaoting Fu, Dr. Ali Luo, and Prof. Biwei Jiang for their helpful discussions. We gratefully acknowledge the anonymous referee for improvement of this letter. 

Funding for SDSS-III has been provided by the Alfred P. Sloan Foundation, the Participating Institutions, the National Science Foundation, and the U.S. Department of Energy Office of Science. The SDSS-III Web site is http://www.sdss3.org/.

SDSS-III is managed by the Astrophysical Research Consortium for the Participating Institutions of the SDSS-III Collaboration including the University of Arizona, the Brazilian Participation Group, Brookhaven National Laboratory, University of Cambridge, Carnegie Mellon University, University of Florida, the French Participation Group, the German Participation Group, Harvard University, the Instituto de Astrofisica de Canarias, the Michigan State/Notre Dame/JINA Participation Group, Johns Hopkins University, Lawrence Berkeley National Laboratory, Max Planck Institute for Astrophysics, Max Planck Institute for Extraterrestrial Physics, New Mexico State University, New York University, Ohio State University, Pennsylvania State University, University of Portsmouth, Princeton University, the Spanish Participation Group, University of Tokyo, University of Utah, Vanderbilt University, University of Virginia, University of Washington, and Yale University.

Guoshoujing Telescope (the Large Sky Area Multi-Object Fiber Spectroscopic Telescope LAMOST) is a National Major Scientific Project built by the Chinese Academy of Sciences. Funding for the project has been provided by the National Development and Reform Commission. LAMOST is operated and managed by the National Astronomical Observatories, Chinese Academy of Sciences.

{\it Facilities:} \facility{SDSS}, \facility{LAMOST}

\end{document}